\documentclass[UKenglish]{lipics}

\newcommand{\ie}{\textit{i.e., }}

\theoremstyle{plain}
\newtheorem{proposition}[theorem]{Proposition}

\title{Non-termination of Dalvik bytecode \emph{via} compilation to CLP}
\titlerunning{Non-termination of Dalvik bytecode}
\author{\'Etienne Payet}
\author{Fred Mesnard}
\affil{Universit\'e de La R\'eunion, EA2525-LIM\\
  Saint-Denis de La R\'eunion, F-97490, France\\
  \texttt{\{etienne.payet,frederic.mesnard\}@univ-reunion.fr}}
\authorrunning{\'E. Payet and F. Mesnard}

\def\copyrightline{}
\Copyright{\'Etienne Payet and Fred Mesnard}



\begin{document}

\maketitle

\begin{abstract}
  We present a set of rules for compiling a Dalvik bytecode
  program into a logic program with array constraints.
  Non-termination of the resulting program entails that of the
  original one, hence the techniques we have presented
  before for proving non-termination of constraint logic
  programs can be used for proving non-termination of 
  Dalvik programs.
\end{abstract}

%%%%%%%%%%%%%%%%%%%%%%%%
\section{Introduction}
%%%%%%%%%%%%%%%%%%%%%%%%
Android is currently the most widespread operating system for mobile
devices. Applications running on this system can be downloaded from
anywhere, hence reliability is a major concern for its users.
In this paper, we consider applications that may run into an infinite
loop, which may cause a resource exhaustion, for instance the battery
if the loop continuously uses a sensor as the GPS.
Android programs are written in Java and compiled to the
Google's Dalvik Virtual Machine (DVM) bytecode
format~\cite{dalvik-doc} before installation on a device.
We provide a set of rules for compiling a Dalvik bytecode program into
a constraint logic program~\cite{Jaffar98a}. 
Non-termination of the resulting program
entails that of the original one, hence the technique
we have presented before~\cite{PayetM09} for proving % existential
non-termination of constraint logic programs can be used for proving
non-termination of Dalvik programs.
We model the memory and the objects it contains with \emph{arrays},
so we compile Dalvik programs to logic programs with array constraints
and we consider the theory of arrays presented in~\cite{BraleyMS06}.

%%%%%%%%%%%%%%%%%%%%%%%%%%%%%%%%%%%%%%%
\section{The Dalvik Virtual Machine}
%%%%%%%%%%%%%%%%%%%%%%%%%%%%%%%%%%%%%%%
We briefly describe the operational semantics of the DVM
(see~\cite{dalvik-doc} for a complete description).
Unlike the JVM which is stack-based, the DVM is
register-based. Each method uses its own array of registers
and invoked methods do not affect the registers of invoking methods.
The number of registers used by a method is statically known.
At the beginning of an execution, the $N$ arguments to a method
land in its last $N$ registers and the other registers are
initialized to 0. Many Dalvik bytecode instructions are
similar, so we concentrate on a restricted set
which exemplifies the operations that the DVM performs. 
\begin{itemize}
\item $\mathit{const}\ d,c$ Move constant $c$ into register $d$
  (\ie the register at index $d$ in the array of registers of the method
  where this instruction occurs).
\item $\mathit{move}\ d,s$ Move the content of register $s$
  into register $d$.
\item $\mathit{add}\ d,s,c$ Store the sum of the content
  of register $s$ and constant $c$ into register $d$.
\item $\mathit{if\text{-}lt}\ i,j,q$ If the content of register $i$
  is less than the content of register $j$ then jump to program point $q$,
  otherwise execute the immediately following instruction.
\item $\mathit{goto}\ q$ Jump to program point $q$.
\item $\mathit{invoke}\ S, m$
  where $S=s_0,s_1,\ldots,s_p$ is a sequence of register indexes
  and $m$ is a method.
  The content $r^{s_0}$ of register $s_0$, \ldots,
  $r^{s_p}$ of register $s_p$ are the \emph{actual parameters}
  of the call. Value $r^{s_0}$ is called \emph{receiver} of the call
  and must be $0$ (the equivalent of {\tt{}null} in Java) or a reference to an
  object $o$. In the former case, the computation stops with an exception.
  Otherwise, a lookup procedure is started from the class of $o$ upwards along
  the superclass chain, looking for a method with the same signature as $m$.
  That method is run from a state where its last registers are bound to
  $r^{s_0},r^{s_1},\ldots,r^{s_p}$.
\item $\mathit{return}$ Return from a void method.
% \item $\mathit{return}\ s$ Return from a non-void method with the content of
%   register $s$ as result.
% \item $\mathit{move\text{-}result}\ d$ Move the result of the most recent called
%   method into register $d$. This instruction must immediately
%   follow an $\mathit{invoke}$ instruction.
\item $\mathit{new\text{-}instance}\ d,\kappa$ Move a reference to a new object
  of class $\kappa$ % (which is properly initialised)
  into register $d$.
\item $\mathit{iget}\ d,i,f$ (resp. $\mathit{iput}\ s,i,f$)
  The content $r^i$ of register $i$ must be $0$
  or a reference to an object $o$. If $r^i$ is $0$, the computation 
  stops with an exception. Otherwise, $o(f)$ (the value of field $f$ of $o$)
  is stored into register $d$
  (resp. the content of register $s$ is stored into $o(f)$).
\end{itemize}

%%%%%%%%%%%%%%%%%%%%%%%%%%%%%%%%%%%%%%%%%
\section{Compilation to CLP clauses}
%%%%%%%%%%%%%%%%%%%%%%%%%%%%%%%%%%%%%%%%%
%
We model a \emph{memory} as a pair $(a,i)$ where $a$ is an array of
\emph{objects} and $i$ is the index into this array where the next
insertion will take place.
An \emph{object} $o$ is an array of terms of the form 
$[w,f_1(v_1),\ldots,f_n(v_n)]$ where $w$ is the name of the
class of $o$, $f_1$, \dots, $f_n$ are the names of the fields
defined in this class and $v_1$, \dots, $v_n$ are the current
values of these fields in $o$. So, the first component of a memory
is an array of arrays of terms and a memory location is an index
into this array. Memory locations start at 1 and 0
corresponds to the \verb+null+ value.

Our compilation rules are given in
Fig.~\ref{fig:compilation-simple}--\ref{fig:compilation-memory-constraints}.
We associate a predicate symbol $p_q$ to each program point $q$
of the Dalvik program $P$ under consideration.
We generate clauses with constraints on integer and array terms.
Our constraint theory combines the theory of integers with that of
arrays defined in~\cite{BraleyMS06}.
Our CLP domain of computation $\mathcal{D}$ (values interpreting constraints)
is the union of $\mathbb{Z}$ with the set $\mathit{Obj}$ of
arrays of terms of the form $f(i)$ where $i$ is an integer
and with the set of arrays of elements of $\mathit{Obj}$.
The read $a[i]$ returns the value stored at position $i$ of the array $a$
and the write $a\{i\leftarrow e\}$ is $a$ modified so that position $i$
has value $e$. For multidimensional arrays, we abbreviate $a[i]\cdots[j]$
with $a[i,\ldots,j]$.

Each rule considers an instruction $\mathit{ins}$ occurring at a program
point $q$.
% We let $q+1$ denote the program point where the instruction
% immediately following $\mathit{ins}$ in the code occurs. 
%
We let $\tilde{V} = V_0,\dots,V_{r-1}$ and $\tilde{V}' = V'_0,\dots,V'_{r-1}$
be sequences of distinct variables where $r$ is the number of registers
used by the method where $\mathit{ins}$ occurs. For each $i\in[0,r-1]$,
variable $V_i$ (resp. $V'_i$) models the content of register $i$ before
(resp. after) executing $\mathit{ins}$.
We let $M$ denote the input memory and $M'$ the output memory.
So, $\tilde{V}$ and $M$ (or $[A,I]$) in the head of the
clauses are input parameters while $M'$ is an output parameter.
%
% We use variable $R$ for storing the value returned by a method.
%
We let $\mathit{id}$ denote the sequence
$(V'_0=V_0,\dots,V'_{r-1}=V_{r-1})$ and $\mathit{id}_{-i}$
(where $i\in[0,r-1]$) the sequence
$(V'_0=V_0,\dots,V'_{i-1}=V_{i-1},V'_{i+1}=V_{i+1}\dots,V'_{r-1}=V_{r-1})$.
By $|\tilde{X}|$ we mean the length of sequence $\tilde{X}$.
For any method $m$, $q_m$ is the program point where 
$m$ starts,
$\mathit{reg}(m)$ is the number of registers used by $m$
and $\mathit{sign}(m)$ is the set of all the methods
with the same signature as $m$.

Some compilation rules are rather straightforward.
For instance, $\mathit{const}\ d,c$ moves constant $c$ into register $d$,
so in Fig.~\ref{fig:compilation-simple} the output register variable $V'_d$
is set to $c$ while the other register variables remain unchanged
(modelled with $\mathit{id}_{-d}$). Rules for $\mathit{move}$,
$\mathit{add}$ and $\mathit{goto}$ are similar.
In Fig.~\ref{fig:compilation-invoke}, we consider method calls. The instruction
$\mathit{invoke}\ s_0,\dots,s_p,m$ is compiled into a set of clauses
(one for each method with the same signature as $m$) which
impose that $V_{s_0}$ (the receiver of the call)
is a non-\verb+null+ location (\ie $V_{s_0}>0$). Therefore,
if $V_{s_0}\leq 0$, the execution of the generated CLP program
fails, as the original Dalvik program. If $V_{s_0}>0$, the
lookup procedure begins.
For each $m'\in\mathit{sign}(m)$, this is modelled with the call
$\mathit{lookup}_P(M, V_{s_0}, m, q_{m'})$ which starts from the class
of the object at location $V_{s_0}$ in memory $M$ and searches for
the closest method $m''$ with the same signature as $m$ upwards
along the superclass chain. If $m''=m'$, this call succeeds, otherwise
it fails. Then, $m'$ is executed, modelled with
$p_{q_{m'}}(\tilde{X}_{m'},M,M_1)$, with some registers $\tilde{X}_{m'}$
initialized as expected.
% (\ie the arguments of the calls stand in the 
% last registers and the other registers are set to $0$).
When the execution of $m'$ has finished, control jumps to the following
instruction (\ie $p_{q+1}(\tilde{V}',M_1,M')$).
% Instruction $\mathit{return}$ stops the execution of the current method,
% hence the body of the generated clause contains no call.
% Instruction $\mathit{return}\ s$ returns the content of
% register $s$ into variable $R$ and stops the
% execution of the current method (hence, the body of the
% generated clause contains no call).
% Instruction $\mathit{move\text{-}result}\ d$ moves 
% the result of the most recent called method, which is
% stored in $R$, into register $d$.
In Fig.~\ref{fig:compilation-memory-constraints}, we consider
some memory-related instructions that we compile to clauses
with array constraints.

%%% FIGURE: compilation of some simple instructions.
\begin{figure}%[p]
  \begin{subequations}
    \begin{equation}\label{eq:const}
      \frac{\mathit{const}\ d,c}
      {p_q(\tilde{V},M,M') \leftarrow
        \{V'_d=c\} \cup \mathit{id}_{-d},\
        p_{q+1}(\tilde{V}',M,M')
      }
    \end{equation}
    % 
%     \begin{equation}\label{eq:move}
%       \frac{\mathit{move}\ d,s}
%       {p_q(\tilde{V},M,M',R) \leftarrow
%         \{V'_d=V_s\}\cup \mathit{id}_{-d},\
%         p_{q+1}(\tilde{V}',M,M',R)
%       }
%     \end{equation}
%     
%     \begin{equation}\label{eq:add}
%       \frac{\mathit{add}\ d,s_1,s_2}
%       {p_q(\tilde{V},M,M',R) \leftarrow
%         \{V'_d=V_{s_1}+V_{s_2}\}\cup \mathit{id}_{-d},\
%         p_{q+1}(\tilde{V}',M,M',R)
%       }
%     \end{equation}
    % 
    \begin{equation}\label{eq:if-eqz}
      \frac{\mathit{if\text{-}lt}\ i,j,q'}
      {\begin{array}{lll}
          \{ & p_q(\tilde{V},M,M') \leftarrow
          \{V_i<V_j\}\cup\mathit{id},\
          p_{q'}(\tilde{V}',M,M'), & \\
          & p_q(\tilde{V},M,M') \leftarrow
          \{V_i\geq V_j\}\cup\mathit{id},\
          p_{q+1}(\tilde{V}',M,M') & \}
        \end{array}
      }
    \end{equation}
    % 
%     \begin{equation}\label{eq:goto}
%       \frac{\mathit{goto}\ q'}
%       {p_q(\tilde{V},M,M',R) \leftarrow
%         \mathit{id},\ p_{q'}(\tilde{V}',M,M',R)
%       }
%     \end{equation}
%     
  \end{subequations}
  \caption{Compilation of some simple Dalvik instructions.}
  \label{fig:compilation-simple}
\end{figure}

%%% FIGURE: instructions related to method calls.
\begin{figure}%[p]
  \begin{subequations}
    \begin{equation}\label{eq:invoke}
      \frac{\mathit{invoke}\ s_0,\dots,s_p,m}
      {
        \left\{
          \begin{array}{r@{\hskip 1mm}c@{\hskip 1mm}l|}
            p_q(\tilde{V},M,M') & \leftarrow & \{V_{s_0}>0\} \cup \mathit{id},\\
            & & \mathit{lookup}_P(M, V_{s_0}, m, q_{m'}),\\
            & & p_{q_{m'}}(\tilde{X}_{m'},M,M_1),\\
            & & p_{q+1}(\tilde{V}',M_1,M') 
          \end{array}
          \begin{array}{l}
            m'\in \mathit{sign}(m)\\
            \text{and } \tilde{X}_{m'}=0,\dots,0,V_{s_0},\dots,V_{s_p}\\
            \text{with } |\tilde{X}_{m'}| = \mathit{reg}(m')
          \end{array}
        \right\}
      }
    \end{equation}
    \begin{equation}\label{eq:return}
      \frac{\mathit{return}}{p_q(\tilde{V},M,M') \leftarrow \{M'=M\}}
    \end{equation}
%
%     \begin{equation}\label{eq:return}
%       \frac{\mathit{return}\ s}{p_q(\tilde{V},M,M',R) \leftarrow \{M'=M, R=V_s\}}
%     \end{equation}
    % 
%     \begin{equation}\label{eq:move-result}
%       \frac{\mathit{move\text{-}result}\ d}
%       {p_q(\tilde{V},M,M',R) \leftarrow
%         \{V'_d=R\}\cup\mathit{id}_{-d},\
%         p_{q+1}(\tilde{V}',M,M',R)}
%     \end{equation}
  \end{subequations}
  \caption{Compilation of some Dalvik instructions related to method calls.}
  \label{fig:compilation-invoke}
\end{figure}

%%% FIGURE: compilation of memory-related instructions, where the memory
%%% is handled with constraints
\begin{figure}%[p]
  \begin{subequations}
    \begin{equation}\label{eq:newinstance-constraints}
      \frac{\begin{array}{c}
          \mathit{new\text{-}instance}\ d,\kappa\\
          \text{$w$ is the name of class $\kappa$ and
            $f_1,\dots,f_n$ are the names of the fields defined in $\kappa$}
        \end{array}
      }
      {\begin{array}{l}%{r@{\hskip 1mm}c@{\hskip 1mm}l}
          p_q(\tilde{V},[A,I],M') \leftarrow
          \big\{O[0]=w,\ O[1]=f_1(0),\ \ldots,\ O[n]=f_n(0),\\[1ex]
          \hspace{8mm}A_1=A\{I\leftarrow O\},\ V'_d=I,\
          I_1 = I + 1\big\}\cup \mathit{id}_{-d},\
          p_{q+1}(\tilde{V}',[A_1,I_1],M')
        \end{array}
      }
    \end{equation}
    \begin{equation}\label{eq:iget-constraints}
      \frac{\mathit{iget}\ d,i,f}
      {p_q(\tilde{V},[A,I],M') \leftarrow
        \big\{V_i>0,\ A[V_i,F] = f(V'_d)\big\} \cup \mathit{id}_{-d},\
        p_{q+1}(\tilde{V}',[A,I],M')
      }
    \end{equation}
    \begin{equation}\label{eq:iput-constraints}
      \frac{\mathit{iput}\ s,i,f}
      {\begin{array}{l}%{r@{\hskip 1mm}c@{\hskip 1mm}l}%{rcl}
          p_q(\tilde{V},[A,I],M') \leftarrow
          \big\{V_i>0,\ O=A[V_i],\ O[F] = f(X),\
          O_1 = O\{F\leftarrow f(V_s)\},\\[1ex]
          \hspace{8mm}A_1 = A\{V_i \leftarrow O_1\}\big\} \cup \mathit{id},\
          p_{q+1}(\tilde{V}',[A_1,I],M')
        \end{array}
      }
    \end{equation}
  \end{subequations}
  \caption{Compilation of some memory-related instructions.}
  \label{fig:compilation-memory-constraints}
\end{figure}

\begin{theorem}
  \label{theorem:underapproximation}
  Let $P$ be a Dalvik bytecode program and $P_{\mathit{CLP}}$ its
  CLP compilation. If there is a computation 
  $p_{q_0}p_{q_1}\ldots$
  in $P_{\mathit{CLP}}$ then there is an execution
  $q_0q_1\ldots$ of $P$.
\end{theorem}
More precisely, if there is a finite (resp. infinite) computation in
$P_{\mathit{CLP}}$ starting from a query $p_{q_0}(\tilde{v},[a,i],M')$ (where
$\tilde{v}$, $a$ and $i$ are values in $\mathcal{D}$
and $M'$ is an output variable), then there is a finite
(resp. infinite) execution of $P$, using the same program points,
starting from values corresponding to $\tilde{v}$ and $a$ in the DVM
registers and memory.

%%%%%%%%%%%%%%%%%%%%%%%%%%%%%%%%%%%%%%%
\section{Non-termination inference}
%%%%%%%%%%%%%%%%%%%%%%%%%%%%%%%%%%%%%%%
The following proposition is a CLP reformulation of a
result presented in~\cite{Gupta08}.
\begin{proposition}\label{proposition:non-termination}
  Let
  $r = p(\tilde{x}) \leftarrow c, p(\tilde{y})$
  and
  $r' = p'(\tilde{x}') \leftarrow c', p(\tilde{y}')$
  be some clauses.
%  where $\tilde{x}$, $\tilde{y}$, $\tilde{x}'$, $\tilde{y}'$
%  are disjoint sequences of distinct variables and
%  $c$ and $c'$ are satisfiable constraints.
%  on variables $\tilde{x},\tilde{y}$ and $\tilde{x}',\tilde{y}'$,
%  respectively.
  Suppose there exists a set $\mathcal{G}$ such that formul\ae{}
  $\big[\forall \tilde{x} \exists\tilde{y} \ \tilde{x}\in\mathcal{G}
  \Rightarrow (c \land \tilde{y}\in\mathcal{G})\big]$
  and
  $\big[\exists \tilde{x}'\exists \tilde{y}'\
  c' \land \tilde{y}'\in\mathcal{G}\big]$
  are true. Then, $p'$ has an infinite computation in $\{r,r'\}$.
\end{proposition}

Consider the Android program in Fig.~\ref{fig:alias-loop}, with
the Java syntax on the left and the corresponding Dalvik bytecode
$P$ on the right, where \verb+v0+, \verb+v1+, \ldots{} denote
registers 0, 1,~\ldots{}
Method \verb+loop+ in class \verb+MyActivity+ is called
when the user taps a button displayed by the application.
Execution of this method does not terminate because in the
call to \verb+m+, the objects \verb+o1+ and \verb+o2+ are
aliased and therefore by decrementing \verb+x.i+ we are 
also decrementing \verb+this.i+ in the loop of method \verb+m+.
\begin{figure}
  \footnotesize
\begin{verbatim}
public class Loops {                  .method public m(ILoops)V
  int i;                                  .registers 4
  public void m(int n, Loops x) {     0:  iget v0, v1, Loops->i:I
    while (this.i < n) {              1:  if-lt v0, v2, 3
      this.i++;                       2:  return-void
      x.i--;                          3:  iget v0, v1, Loops->i:I
    }                                 4:  add-int/lit8 v0, v0, 0x1
  }                                   5:  iput v0, v1, Loops->i:I
}                                     6:  iget v0, v3, Loops->i:I
                                      7:  add-int/lit8 v0, v0, -0x1
                                      8:  iput v0, v3, Loops->i:I
                                      9:  goto 0
                                      .end method

public class MyActivity extends Activity {
  ...                                 .method public loop(Landroid/view/View;)V
  public void loop(View v) {              .registers 5
    Loops o1 = new Loops();           10:  new-instance v0, Loops     
    Loops o2 = o1;                    11:  invoke-direct {v0}, Loops-><init>()V
    o1.m(2, o2);                      12:  move-object v1, v0
  }                                   13:  const/16 v2, 0x2
  ...                                 14:  invoke-virtual {v0, v2, v1}, Loops->m(ILoops)V
}                                     15:  return-void
                                      .end method
\end{verbatim}
  \caption{The non-terminating method \texttt{loop} is called
    when the user taps a button.}
  \label{fig:alias-loop}
\end{figure}
We get the following clauses for program points 0 and 14:
\[\begin{array}{r@{\hskip 1mm}c@{\hskip 1mm}l}
  p_0(\tilde{V},[A,I],M') & \leftarrow &
  \{A[V_1,F] = i(V'_0)\}\cup\mathit{id}_{-0},\
  p_1(\tilde{V}',[A,I],M')\\[2ex]
  p_{14}(\tilde{V},M,M') & \leftarrow &
  \{V_0>0\}\cup\mathit{id},\
  \mathit{lookup}_P(M,V_0,\texttt{Loops->m(ILoops)V},0),\\
  & & p_0(0,V_0,V_2,V_1,M,M_1),\
  p_{15}(\tilde{V}',M_1,M')
\end{array}\]

Let $P_{\mathit{CLP}}$ denote the CLP program resulting from the compilation 
of $P$. The set of \emph{binary unfoldings}~\cite{Codish99a} of $P_{\mathit{CLP}}$
contains the following clauses
\[\begin{array}{cl}%{cr@{\hskip 1mm}c@{\hskip 1mm}l}
  r: & 
  p_0(\tilde{V},[A,I],M') \leftarrow
  \big\{
  V_1>0,\ O = A[V_1],\ O[F]=i(X),\ X < V_2,\\[1ex]
  & \hspace{0.8cm}O_1=O\{F\leftarrow i(X+1)\},\ A_1=A\{V_1\leftarrow O_1\},\\[1ex]
  & \hspace{0.8cm}V_3>0,\ O' = A_1[V_3],\ O'[F']=i(X'),\ V'_0 = X'-1,\\[1ex]
  & \hspace{0.8cm}O'_1=O'\{F'\leftarrow i(V'_0)\},\ A_2=A_1\{V_3\leftarrow O'_1\} % \\[1ex]
  % &
  \big\} \cup\mathit{id}_{-0},\
  p_0(\tilde{V}',[A_2,I],M') \\[2ex]
  r': &
  p_{10}(\tilde{V},[A,I],M') \leftarrow
  \{O[0]=\mathit{loops},\ O[1] = i(0),\
  A_1=A\{I\leftarrow O\},\\[1ex]
  & \hspace{0.8cm}I_1=I+1,\ I > 0\},\
  p_0(0,I,2,I,[A_1,I_1],M_1)
\end{array}\]
where $r$ corresponds to the path
$0 \rightarrow 1 \rightarrow 3 \rightarrow 4 \rightarrow \cdots \rightarrow 9 \rightarrow 0$
and $r'$ to the path
$10 \rightarrow 11 \rightarrow 12 \rightarrow 13 \rightarrow 14 \rightarrow 0$
in $P$.
In $r'$, $O$ corresponds to both $o_1$ and $o_2$, which expresses 
that $o_1$ and $o_2$ are aliased. Note that $I$, the address of $O$, is passed
to $p_0$ both as second and fourth parameter, which corresponds in $r$ to $V_1$
(\verb+this+ in method \verb+m+) and $V_3$ (\verb+x+ in \verb+m+).
Moreover, when $V_1=V_3$ in $r$, we have $O'=O_1$, $F'=F$ and $X'=X+1$, hence
$V'_0=X'-1=X$. Therefore, we have $O'_1=O$, so $A_2=A$.
The logical formul\ae{} of Proposition~\ref{proposition:non-termination}
are true for the set
$\mathcal{G}=\{(\tilde{v},\mathit{mem},\mathit{mem}') \in \mathcal{D}^3|v_1=v_3\}$.
Hence, $p_{10}$ has an infinite computation in $\{r,r'\}$, which
implies~\cite{Codish99a} that $p_{10}$ has an infinite computation
in $P_{\mathit{CLP}}$.
So by Theorem~\ref{theorem:underapproximation}, $P$
has an infinite execution from program point~10.

%%%%%%%%%%%%%%%%%%%%%%%%
\section{Future Work}
%%%%%%%%%%%%%%%%%%%%%%%%
We plan to implement the technique described above and
to write a solver for array constraints.
Currently, our compilation rules only consider the operational
semantics of Dalvik, a part of the Android platform. We also plan
to extend them by considering the operational semantics of other 
components of Android, for instance
\emph{activities} that we have studied in~\cite{PayetS14}.

\bibliographystyle{plain}
% \bibliography{wst14}

\end{document}